\documentclass[final]{siamltex}

\usepackage{latexsym,amsfonts,amsmath,amssymb,enumerate}
\usepackage{xspace}
\usepackage[margin=1in]{geometry}

\newtheorem{observation}[theorem]{Observation}

\def\expected{\ensuremath{\mathbb{E}}}

\newcommand{\hide}[1]{}

\def\var{\mathrm{var}}

\newcommand{\Emptynum}{\ensuremath{n_0}\xspace}

\newcommand{\M}{\xi}

\title{\Large Distributed Selfish Load Balancing\thanks{%
  Preliminary version in proceedings of ACM-SIAM Symposium on Discrete
  Algorithms (SODA) 2006.
  This work was partially supported by the EPSRC grants ``Discontinuous
  Behaviour in the Complexity of Randomized Algorithms'' and
  ``Algorithmics of Network-sharing Games'', and by
  the Natural Sciences and Engineering Research Council of Canada
 (NSERC) discovery grant 250284-2002.}}

\author{
  Petra Berenbrink\footnotemark[2] \and
  Tom Friedetzky\footnotemark[3] \and
  Leslie Ann Goldberg\footnotemark[4] \and\newline
  Paul W. Goldberg\footnotemark[4] \and
  Zengjian Hu\footnotemark[2] \and
  Russell Martin\footnotemark[4]
}

\date{}

\begin{document}
\maketitle

\renewcommand{\thefootnote}{\fnsymbol{footnote}}
\footnotetext[2]{
  School of Computing Science, Simon Fraser University, Canada
}
\footnotetext[3]{
  Department of Computer Science, University of Durham, U.K.
}
\footnotetext[4]{
  Department of Computer Science, University of Liverpool, U.K.
}
\renewcommand{\thefootnote}{\arabic{footnote}}

\begin{abstract}
Suppose that a set of $m$ tasks are to be shared as equally as
possible amongst a set of $n$ resources. A game-theoretic mechanism to
find a suitable allocation is to associate each task with a ``selfish
agent'', and require each agent to select a resource, with the cost of a
resource being the number of agents to select it. Agents would then be
expected to migrate from overloaded to underloaded resources, until
the allocation becomes balanced.

Recent work has studied the question of how this can take place within
a distributed setting in which agents migrate selfishly without any
centralized control. In this paper we discuss a natural protocol for
the agents which combines the following desirable features: It can be
implemented in a strongly distributed setting, uses no central
control, and has good convergence properties.  For $m\gg n$, the
system becomes approximately balanced (an $\epsilon$-Nash equilibrium)
in expected time $O(\log\log m)$.  We show using a martingale
technique that the process converges to a perfectly balanced
allocation in expected time $O(\log\log m+n^4)$. We also give a lower
bound of $\Omega(\max\{\log\log m,n\})$ for the convergence time.
\end{abstract}

\section{Introduction}\label{sec:intro}

Suppose that a consumer learns the price she would be charged by some
domestic power supplier other than the one she is currently using. It
is plausible that if the alternative price is lower than the price she
is currently paying, then there is some possibility that she will
switch to the new power supplier. Furthermore, she is more likely to
switch if the ratio of current price to new price is large. If there
is only a small saving, then it becomes unattractive to make the
switch, since an influx of new business (oneself and other consumers)
may drive up the price of the new power supplier and make it no longer
competitive.

We study a simple mathematical model of the above natural rule, in
the context of a load balancing (or task allocation) scenario that
has received a lot of recent attention.  We assume the presence of
many individual users who may assign their tasks to chosen
resources.  The users are {\em selfish} in the sense that they
attempt to optimize their own situation, i.e., try to assign their
tasks to minimally loaded resources, without trying to optimize the
global situation.  In general, a {\em Nash equilibrium} among a set
of selfish users is a state in which no user has the incentive to
change her current decision.  In our setting, this corresponds to no
user having an incentive to reallocate their task to some other
resource. An $\epsilon$-Nash equilibrium is a standard notion of an
approximate Nash equilibrium, and is a state in which no user can
reduce her cost by a multiplicative factor of less than $1-\epsilon$
by changing action.  Here we do not focus on the {\em quality} of
equilibria, but rather on the (perhaps more algorithmic) question of
convergence time to such a state.

We assume a strongly distributed and concurrent setting, i.e., there
is no centralized control mechanism whatsoever, and all users may
choose to reallocate their tasks at the same time.  Thus, we do not
(and cannot) use the {\em Elementary Step System}~\cite{ors}
(discussed in more detail in the next section), where the assumption
is that at most one user may reallocate her task at any given stage.

Throughout we let $m$ denote the number of tasks (in the above
discussion, customers) and $n$ the number of resources (power
suppliers). As hinted in the above discussion, we assume that
typically $m\gg n$. In a single time step (or round) each task does
the following.  Let $i$ be the resource currently being used by the
task.  Select $j$ uniformly at random from $\{1,\ldots,n\}$ and find
the load of resource $j$. Let $X_i$ and $X_j$ be the loads of
resources $i$ and $j$ respectively. If $X_j<X_i$, migrate from $i$ to
$j$ with a probability of $1-X_j/X_i$; the transition from round $t$
to round $t+1$ is given in Figure \ref{fig:protslow}.  Notice that if
we had unconditional migrations, i.e., without an additional coin flip
(move only with probability $1 - X_j(t) / X_i(t)$), then this may lead
to an unstable system; consider for example the case $m=2$ with
initially most tasks assigned to one of the resources: the overload
would oscillate between the two resources, with a load ratio tending
towards 2:1. (This observation about the risk of oscillation has also
been made in similar contexts in~\cite{FV05, FRV06}, and we will not
elaborate on it further.)

\begin{figure}[ht]
  \begin{center}
  \framebox{
  \begin{minipage}{\linewidth}
    \begin{tabbing}\hspace*{1em}\=\hspace*{1em}\=\kill
    {\bf For} each task $b$ {\bf do} in parallel \\
    \> Let $i_b$ be the current resource of task $b$ \\
    \> Choose resource $j_b$ uniformly at random \\
    \> Let $X_{i_b}(t)$ be the current load of resource $i$  \\
    \> Let $X_{j_b}(t)$ be the current load of resource $j$\\
    \> {\bf If} $X_{i_b}(t) > X_{j_b}(t)$ {\bf then} \\
    \> \>  Move task $b$ from resource $i_b$ to $j_b$ with
           probability $1 - X_{j_b}(t) / X_{i_b}(t)$
    \end{tabbing}
  \end{minipage}
  }
  \end{center}
  \caption{The protocol with ``neutral moves'' allowed.}
  \label{fig:protslow}
\end{figure}

It can easily be seen that, if all tasks use the above policy, then the
expected load of every resource at the next step is $m/n$:

\begin{observation}
\label{obs:m/n}
Regardless of the load distribution at time step $t$, the expected load
of every resource at the next step is $m/n$.
\end{observation}

\begin{proof}
To see this, assume that the loads $X_i(t)$ are arranged in descending
order so that $X_j(t)\geq X_{j+1}(t)$ and note that
\[
\expected[X_i(t+1)]=X_i(t)
+\sum_{\ell=1}^{i-1} \frac{1}{n}X_\ell(t)\Bigl( 1-\frac{X_i(t)}{X_\ell(t)} \Bigr)
-\sum_{\ell=i+1}^{n} \frac{1}{n}X_i(t)\Bigl( 1-\frac{X_\ell(t)}{X_i(t)} \Bigr)
\]
\[
=X_i(t)
+\frac{1}{n}\sum_{\ell=1}^{i-1} (X_\ell(t) - X_i(t))
-\frac{1}{n}\sum_{\ell=i+1}^{n} (X_i(t) - X_\ell(t))
\]
\[
=X_i(t)+\frac{1}{n}\sum_{\ell=1}^{n} (X_\ell(t) - X_i(t))
=\frac{1}{n}\sum_{\ell=1}^{n}X_\ell(t) = \frac{m}{n}.
\]
\end{proof}

This provides a compelling motivation for the policy, which is that as
a result, no task has an incentive to deviate unilaterally from this
policy.  This implies that in the terminology of~\cite{em05} it is a
{\em Nash rerouting policy}. It is also a simple {regret-minimizing}
policy in the sense of~\cite{bedl} since the average cost of resources
used by an agent is no higher than the best choice of a single
resource to be used repeatedly.  Although the above rule is very
natural and has the nice properties described above, we show that it
may take a long time to converge to a perfectly balanced allocation of
tasks to resources. We address this problem as follows.  Define a {\em
neutral move} to be a task migration from a resource with load $\ell$
at time $t$ to a resource with load $\ell-1$ at time $t$ (so, if no
other task migrates, then the cost to the migrating task is
unchanged.)  We consider a modification in which neutral moves are
specifically disallowed (see Figure~\ref{fig:prot}).  That
seemingly-minor change ensures fast convergence from an almost
balanced state to a perfectly-balanced state.  To summarize, here are
the most important features of the modified protocol:

\begin{itemize}\itemsep0ex

\item We do not need any global information whatsoever (apart from the
number of available resources); in particular, a task does not need to
know the total number of tasks in the system.  Also, it is strongly
distributed and concurrent. If additional tasks were to enter the
system, it would rapidly converge once again, with no outside
intervention.

\item A migrating task needs to query the load of only one other
resource (thus, doing a constant amount of work in each round).

\item When a task finds a resource with a significantly smaller load
(that is, a load that is smaller by at least two), the migration
policy is {\it exactly\/} the same as that used by the Nash rerouting
policy of Figure~\ref{fig:protslow}, so the incentive is to use that
probability.

\item When a task finds a resource with a load that is smaller by
exactly one unit, the migration policy is sufficiently close to the
Nash rerouting policy that the difference in expected load is at most
one, and there is little incentive to deviate.

\item The protocol is simple (as well as provably efficient) enough to
convince users to actually stick to it.

\end{itemize}

\subsection{Related Work}\label{sec:related}

We are studying a simple kind of {\em congestion game}. In their
general form, congestion games specify a set of agents, and a set of
resources, and for each agent, a set of allowed strategies, where a
strategy is the selection of a subset of the resources (in this paper,
any singleton subset is allowed). The cost of a resource is a
non-decreasing function of the number of agents using it, and the cost
for an agent is the sum of the costs of resources it uses. A classical
result due to Rosenthal~\cite{rosenthal} is that pure Nash equilibria
(NE) always exist for congestion games, and this is shown by
exhibiting a potential function; they are a type of {\em potential
game}~\cite{ms}.  The potential function also establishes that pure NE
can be found via sequences of ``better-response'' moves, in which
agents repeatedly switch to lower-cost strategies. The potential
function we use later in this paper is the one of~\cite{rosenthal},
modulo a linear re-scaling.

These results do not show how to find Nash equilibrium efficiently,
the problem being that in the worst case, sequences of these
self-improving moves may be exponentially-long.  The following
questions arise: when can NE be found by any efficient algorithm,
and if so, whether it can be found via an algorithm that purports to
be a realistic model of agents' behavior. Regarding the first of these
questions, the answer is no in the general setting (the problem is
PLS-complete for general congestion games~\cite{fpt}, see
also~\cite{arv, chien}).  PLS-completeness (introduced in~\cite{jpy})
is a generally-accepted criterion for intractability of computational
problems in which we seek a local optimum of a given objective
function.

However, due to the basic fact of~\cite{rosenthal, ms} that pure NE
are sure to result from a sufficiently long better-response sequence,
many algorithms for finding them are based on such sequences.  An
important sub-class is the {\em Elementary step system} (ESS),
proposed in Orda et al.~\cite{ors}, which consists of best-response
moves (where a migrating agent switches not to any improved choice,
but to one that is optimal at the time of migration). For matroid
games (a class of congestion games that includes the ones we consider
here), Ackermann et al.~\cite{arv} show that best-response sequences
must have length polynomial in the number of players, resources,
and maximal rank of the matroids.  In this paper we consider the
special case of singleton congestion games (where players' strategies
are always single resources, thus the ranks of the matroids is 1). For
these games, Ieong et al.~\cite{imnss} give polynomial bounds for
best-response and better-response sequences.  Chien and
Sinclair~\cite{chien} study a version of the ESS in the context of
approximate Nash equilibria, and show that in some cases the {\em
$\epsilon$-Nash dynamics} may find an $\epsilon$-NE where finding an
exact NE is PLS-complete. Mirrokni and Vetta~\cite{mv} study the
convergence rate of the ESS to solutions, and the quality of the
approximation after limited iterations.

While best- and better-response dynamics are a plausible model of
selfish behaviour, the associated algorithms typically require that
migrations be done one-by-one, and another common assumption is that
best (not better) responses are always selected.  This means that to
some extent, agents are being assumed to be governed by a centralized
algorithm that finds a NE, and raises the question of what sort of
{\em distributed} algorithms can do so, especially if agents have
limited information about the state of the system (and so may not be
able to find best responses).  That issue is of central importance to
us in this paper.  Goldberg~\cite{goldberg04} studied situations where
simple better-response approaches can be realised as weakly
distributed algorithms (where each agent looks for moves independently
of the others, but it is assumed that moves take place consecutively,
not simultaneously). In a strongly distributed setting (as we study
here), where moves may occur simultaneously, we need to address the
possibility that a change of strategy may increase an agent's cost. It
may happen that after a best response has been identified, it is not
optimal at the time it is executed. Even-Dar and Mansour~\cite{em05}
consider concurrent, independent rerouting decisions where tasks are
allowed to migrate from overloaded to underloaded resources. Their
rerouting process terminates in expected $O(\log\log m+\log n)$ rounds
when the system reaches a Nash equilibrium. Note that their
convergence rate as a function of the number $n$ of resources is
faster than the one we obtain in this paper. The reason is that is
requires agents to have a certain amount of global knowledge.  A task
is required to know whether its resource is overloaded (having
above-average load) and tasks on underloaded resources do not migrate
at all. Our rerouting policy does not require that agents know
anything other that their current resource load, and the load of a
randomly-chosen alternative. Even-Dar and Mansour also present a
general framework that can be used to show a logarithmic convergence
rate for a wide class of rerouting strategies. Our protocol does not
fall into that class, since we do not require migrations to occur only
from overloaded resources. Note that our lower bound is linear in $n$
(thus, more than logarithmic).

Distributed algorithms have been studied in the Wardrop setting (the
limit of infinitely many agents), for which recent work has also
extensively studied the coordination ratio~\cite{RT, Rbook}.  Fischer
et al.~\cite{FRV06} investigate convergence to Wardrop equilibria for
games where agents select paths through a shared network to route
their traffic. (Singleton games correspond to a network of parallel
links.)  Their re-routing strategies are slightly different to ours
--- they assume that in each round, an agent queries a path with
probability proportional to the traffic on that path. Here we assume
paths (individual elements of a set of parallel links) are queried
uniformly at random, so that agents can be assumed to have minimal
knowledge. As in this paper, the probability of switching to a better
path depends on the latency difference, and care has to be taken to
avoid oscillation. Also in the Wardrop setting, Blum et
al.~\cite{bedl} show that approximate NE is the outcome of {\em
regret-minimizing} rerouting strategies, in which an agent's cost,
averaged over time, should approximate the cost of the best individual
link available to that agent.

Certain generalisations of singleton games have also been
considered. These generalisations are not strictly congestion games
according to the standard definition we gave above, but many ideas
carry over. One version introduced by Koutsoupias and
Papadimitriou~\cite{kp} has been studied extensively in different
contexts (for example~\cite{ms01,CV,fkkms02,ckv02,RT}).  In this
generalisation, each task may have a numerical {\em weight} (sometimes
called traffic, or demand) and each resource has a {\em speed} (or
capacity). The cost of using a resource is the total weight of tasks
using it, divided by its speed. Even-Dar et al.~\cite{ekm03} give a
generalized version of the potential function of~\cite{rosenthal} that
applies to these games, and which was subsequently used
in~\cite{goldberg04}. For these games however, it seems harder to find
polynomial-length best-response sequences.  Feldman et
al.~\cite{fglmr} show how a sequence of steps may lead to NE, under
the weaker condition that the maximal cost experienced by agents must
not increase, but individual steps need not necessarily be
``selfish''.  They also note that poorly-chosen better-response moves
may lead to an exponential convergence rate.  Another generalisation
of singleton games is {\em player-specific} cost
functions~\cite{milchtaich}, which allow different agents to have
different cost functions for the same resource. In this setting there
is no potential function and better-response dynamics may cycle,
although it remains the case that pure NE always exist.

Our rerouting strategy is also related to reallocation processes for
balls into bins games. The goal of a balls into bins game is to
allocate $m$ balls as evenly as possible into $n$ bins. It is
well-known that a fairly even distribution can be achieved if every
ball is allowed to randomly choose $d$ bins and then the ball is
allocated to the least loaded amongst the chosen bins (see~\cite{power}
for an overview). Czumaj et al.~\cite{crs03} consider
such an allocation where each ball initially chooses two bins. They
show that, in a polynomial number of steps, the reallocation process
ends up in a state with maximum load at most $\lceil m/n\rceil+1$.
Sanders et al.~\cite{sek00} show that a maximum load of $\lceil
m/n\rceil+1$ is optimal if every ball is restricted to two random
choices.

In conclusion, this paper sits at one end of a spectrum in which we
study a very simple load-balancing game, but we seek solutions in a
very adverse setting in which agents have, at any point in time, a
minimal amount of information about the state of their environment,
and carry out actions simultaneously in a strongly distributed sense.

\subsection{Overview of our results}

Section~\ref{sec:upper} deals with upper bounds on convergence time.
The main result, Theorem~\ref{thm:main}, is that the protocol of
Figure~\ref{fig:prot} converges to a Nash equilibrium within
expected time $O(\log\log m+n^4)$.

The proof of Theorem~\ref{thm:main} shows that the system becomes
{\em approximately} balanced very rapidly. Specifically,
Corollary~\ref{epsilonnash} shows that if $n\leq m^{1/3}$, then for
all $\epsilon$, either version of the distributed protocol (with or
without neutral moves allowed) attains an $\epsilon$-Nash
equilibrium (where all load ratios are within
$[1-\epsilon,1+\epsilon]$; we use $\epsilon$ to denote a multiplicative
factor as in~\cite{chien}) in expected $O(\log\log m)$ rounds.  The
rest of Section~\ref{sec:upper} analyses the protocol of
Figure~\ref{fig:prot}.  It is shown that within an additional
$O(n^4)$ rounds the system becomes optimally balanced.

In Section \ref{sec:lower}, we provide two lower bound results.  The
first one, Theorem \ref{thm:neutral}, shows that the first protocol
(of Figure \ref{fig:protslow}, including moves that do not
necessarily
yield a strict  improvement for an individual task but allow for simply
``neutral'' moves as well, results in exponential (in $n$) expected
convergence time.  Finally, in Theorem \ref{thm:lower-loglog} we provide
a general lower bound (regardless of which of the two protocols is being
used) on the expected convergence time of $\Omega(\log\log m)$.  This
lower bound matches the upper bound as a function of $m$.

\section{Notation}
\label{sec:notation}

There are $m$~tasks and $n$~resources. An assignment of tasks
to resources is represented as a vector $(x_1,\ldots,x_n)$ in which
$x_i$ denotes the number of tasks that are assigned to
resource $i$.  In the remainder of this paper, $[n]$ denotes
$\{1,\ldots,n\}$. The assignment is a Nash
equilibrium if for all $i\in[n]$ and $j\in[n]$,
$|x_i-x_j|\leq 1$. We study a distributed process for
constructing a Nash equilibrium. The states of the process,
$X(0),X(1),\ldots$, are assignments.  The transition from
state~$X(t)=(X_1(t),\ldots,X_n(t))$ to state~$X(t+1)$ is given by the
greedy distributed protocol in Figure~\ref{fig:prot}.

\begin{figure}[ht]
  \begin{center}
  \framebox{
  \begin{minipage}{\linewidth}
    \begin{tabbing}\hspace*{1em}\=\hspace*{1em}\=\kill
    {\bf For} each task $b$ {\bf do} in parallel \\
    \> Let $i_b$ be the current resource of task $b$ \\
    \> Choose resource $j_b$ uniformly at random \\
    \> Let $X_{i_b}(t)$ be the current load of resource $i$  \\
    \> Let $X_{j_b}(t)$ be the current load of resource $j$\\
    \> {\bf If} $X_{i_b}(t) > X_{j_b}(t)+1$ {\bf then} \\
    \> \>  Move task $b$ from resource $i_b$ to $j_b$ with
           probability $1 - X_{j_b}(t) / X_{i_b}(t)$
    \end{tabbing}
  \end{minipage}
  }
  \end{center}
  \caption{The modified protocol, with ``neutral moves'' disallowed.}
  \label{fig:prot}
\end{figure}

Note that if $X(t)$ is a Nash equilibrium, then $X(t+1)=X(t)$ so the
assignment stops changing.  Here is a formal description of the
transition from a state $X(t)=x$.  Independently, for every $i\in[n]$,
let $(Y_{i,1}(x),\ldots,Y_{i,n}(x))$ be a random variable drawn from a
multinomial distribution with the constraint $\sum_{j=1}^n Y_{i,j}(x)
= x_i$.  ($Y_{ij}$ represents the number of migrations from $i$ to $j$
in a round.)  The corresponding probabilities
$(p_{i,1}(x),\ldots,p_{i,n}(x))$ are given by
\[
p_{i,j}(x) = \left\{
\begin{array}{ll}
\frac1n \left(1- \frac{x_j}{x_i}\right) & \mbox{if  $x_i>x_j+1$,}\\
0 & \mbox{if $i\neq j$ but $x_i \leq x_j+1$,}\\
1 - \sum_{j\neq i} p_{i,j}(x) & \mbox{if $i=j.$}
\end{array}
\right.
\]
Then $X_i(t+1)=\sum_{\ell=1}^n Y_{\ell,i}(x)$.

For any assignment~$x=(x_1,\ldots,x_n)$, let $\overline{x}=\frac1n
\sum_{i=1}^n x_i$.  We define the potential function
$\Phi(x)=\sum_{i=1}^n{(x_i-\overline{x})}^2$. Note that
$\Phi(x)=\sum_{i=1}^n x_i^2-n\overline{x}^2$, and that a single
selfish move reduces the potential.

\section{Upper bound on convergence time}\label{sec:upper}

Our main result is the following.
\begin{theorem}
\label{thm:main}
Let $T$ be the number of rounds taken by the protocol of
Figure~\ref{fig:prot} to reach a Nash equilibrium for the
first
time.
Then $\expected[T]=O(\log\log m+n^4)$.
\end{theorem}

The proof of this theorem proceeds as follows.  First
(Lemma~\ref{lemsqrt}) we give an upper bound on $\expected[\Phi(X(t))]$
which implies (Corollary~\ref{cortau}) that there is a $\tau=O(\log\log
m)$ such that, with high probability, $\Phi(X(\tau))=O(n)$.  We also show
(Observation~\ref{obsu} and Corollary~\ref{upperu}) that $\Phi(X(t))$ is a
super-martingale and (Lemma \ref{lemvariance}) that it has enough
variance.  Using these facts, we obtain the upper bound on the convergence
time.

\smallskip

\noindent {\bf Definition:}\quad Let $S_i(x)=\{j\mid x_j<x_i-1\}$.
$S_i(x)$ is the set of resources that are significantly smaller than
resource~$i$ in state~$x$
(in the sense that their loads are at least two tasks smaller than the load of resource~$i$).  Similarly, let $L_i(x)=\{j\mid
x_j>x_i+1\}$ and let $d_i(x) = \frac1n \sum_{j: |x_i-x_j|\leq 1}
(x_i-x_j)$.

\begin{observation}
$\expected[X_i(t+1)\mid X(t)=x] = \overline{x}+d_i(x)$.
\label{obsexp}
\end{observation}

\begin{proof}
\begin{align*}
\expected[X_i(t+1)\mid X(t)=x] &= \sum_{\ell=1}^n
\expected[Y_{\ell,i}(x)]
= \sum_{\ell=1}^n x_\ell p_{\ell,i}(x) \\
&=
\sum_{\ell\in L_i(x)} x_\ell \frac1n \left(1-
\frac{x_i}{x_\ell}\right) + x_i\left(1 - \sum_{j \in S_i(x)} \frac1n
\left(1- \frac{x_j}{x_i}\right) \right)\\
&= x_i+ \frac1n \left(
\sum_{\ell\in L_i(x)} (x_\ell-x_i)
- \sum_{j \in S_i(x)} (x_i-x_j)
\right)
= x_i+ \frac1n
\sum_{\ell\in L_i(x)\cup S_i(x)} (x_\ell-x_i)\\
&=x_i+ \frac1n
\sum_{\ell=1}^n (x_\ell-x_i)-
\frac1n
\sum_{\ell\not\in L_i(x)\cup S_i(x)} (x_\ell-x_i)\\
&= \overline{x}-
\frac1n
\sum_{\ell\not\in L_i(x)\cup S_i(x)} (x_\ell-x_i)\\
&= \overline{x}+
\frac1n
\sum_{\ell\not\in L_i(x)\cup S_i(x)} (x_i-x_\ell).
\end{align*}
\end{proof}

\begin{observation}
\label{obssumsqexp}
$\sum_{i=1}^n {(\expected[X_i(t+1)\mid X(t)=x])}^2 = n \overline{x}^2+\sum_{i=1}^n d_i(x)^2$.
\end{observation}
\begin{proof} Using Observation~\ref{obsexp},
\[
    \sum_{i=1}^n (\expected[X_i(t+1)\mid X(t)=x])^2
  = \sum_{i=1}^n (\overline{x}+d_i(x))^2
  = n \overline{x}^2 + 2 \overline{x}\sum_{i=1}^n d_i(x) +
    \sum_{i=1}^n d_i(x)^2,
\]
and the second term is zero since $d_i(x)=\expected[X_i(t+1)\mid X(t)=x] -
\overline{x}$.
\end{proof}

\begin{observation}
\label{obsvar}
$\var[X_i(t+1)\mid X(t)=x] \leq
\frac1n
\sum_{\ell\in L_i(x)}
 \left( x_\ell - x_i\right)
 + \frac1n
\sum_{j\in S_i(x)}  \left(x_i-{x_j}\right)$.
\end{observation}

\begin{proof}
\begin{align*}
\var(X_i(t+1)\mid X(t)=x)
&= \sum_{\ell=1}^n \var(Y_{\ell,i}(x))
= \sum_{\ell=1}^n x_\ell p_{\ell,i}(x) (1-p_{\ell,i}(x)) \\
&= \sum_{\ell\in L_i(x)} x_\ell \frac1n
\left( 1-\frac{x_i}{x_\ell}\right)
 (1-p_{\ell,i}(x)) + x_i p_{i,i}(x)
  \left(
  \sum_{j\in S_i(x)} \frac1n \left(1-\frac{x_j}{x_i}\right)
  \right)\\
  &=\frac1n
  \sum_{\ell\in L_i(x)} (x_\ell-x_i)
 (1-p_{\ell,i}(x)) +
 p_{i,i}(x)\frac1n
  \sum_{j\in S_i(x)} (x_i-x_j) \\
  &\leq
\frac1n
  \sum_{\ell\in L_i(x)} (x_\ell-x_i)
  +
 \frac1n
  \sum_{j\in S_i(x)} (x_i-x_j).
\end{align*}\end{proof}

\noindent{\bf Definition:}\quad
For any assignment~$x$,
let $s_i(x) = |\{j\mid x_j = x_i-1 \}|$
and $l_i(x) = |\{j\mid x_j = x_i+1 \}|$.
Let $u_1(x) = \sum_{i=1}^n \sum_{j\in[n]:|x_i-x_j|>1}|x_i-x_j|$ and
$u_2(x) = \sum_{i=1}^n {(s_i(x)-l_i(x))}^2$. Let $u(x)=u_1(x)/n+u_2(x)/n^2$.
We will show that $u(x)$ is on upper bound on the expected potential after one step, starting from
state~$x$. The quantity $u_1(x)$ corresponds to the contribution arising from the sum of
the variances of the individual loads and $u_2(x)$ corresponds to the rest.

\begin{observation}
\label{obsu}
$\expected[\Phi(X(t+1))\mid X(t)=x] \leq u(x)$.
\end{observation}
\begin{proof}
\begin{align*}
\expected[\Phi(X(t+1))\mid X(t)=x] + n\overline{x}^2
& =\sum_{i=1}^n \expected[X_i(t+1)^2\mid X(t)=x]\\
& = \sum_{i=1}^n {(\expected[X_i(t+1)\mid X(t)=x])}^2 +
    \sum_{i=1}^n \var(X_i(t+1)\mid X(t)=x).
\end{align*}
Using Observations~\ref{obssumsqexp} and~\ref{obsvar}, this is at most
$ n \overline{x}^2+\sum_{i=1}^n d_i(x)^2 + u_1(x)/n $.
But
$$d_i(x)
= \frac1n \sum_{j: |x_i-x_j|\leq 1} (x_i-x_j) = \frac1n
 (s_i(x)-\ell_i(x)),$$ so the result follows.
\end{proof}

\begin{lemma}
\label{lemsqrt}
$\expected[\Phi(X(t+1))\mid X(t)=x] \leq n + 2n^{1/2} \Phi(x)^{1/2}$.
\end{lemma}
\begin{proof}
In the proof of Observation~\ref{obsu}, we established that
$\expected[\Phi(X(t+1))\mid X(t)=x] \leq \sum_{i=1}^n d_i(x)^2 + u_1(x)/n$.
Upper-bounding $u_1(x)$ and using $d_i(x)\leq 1$,
we have
$$\expected[\Phi(X(t+1))\mid X(t)=x] \leq n + \frac1n
\sum_{i=1}^n \sum_{j=1}^n|x_i-x_j|,$$
and since $|x_i-x_j| \leq |x_i-\overline{x}| + |x_j-\overline{x}|$,
this is at most $n + 2\sum_{i=1}^n |x_i-\overline{x}|$.
By Cauchy-Schwarz, ${(\sum_i |x_i-\overline{x}|\cdot 1)}^2 \leq
\sum_i {|x_i-\overline{x}|}^2 \sum_i 1$
so
$$\expected[\Phi(X(t+1))\mid X(t)=x] \leq n + 2 {(n \sum_{i=1}^n {|x_i-\overline{x}|}^2)}^{1/2}.$$
\end{proof}

\begin{corollary}
\label{corjensen}
$\expected[\Phi(X(t+1))] \leq n + 2n^{1/2}
{(\expected[\Phi(X(t))])}^{1/2}$.
\end{corollary}

\begin{proof} Using Lemma~\ref{lemsqrt},
$\expected[\Phi(X(t+1))] \leq n + 2n^{1/2}
\expected[ f^{1/2} ]$
where $f$ denotes the random variable $\Phi(X(t))$.
By Jensen's inequality
$\expected[f^{1/2}]
\leq {( \expected[f] )}^{1/2}$ since the square-root function is concave, so
we get $\expected[\Phi(X(t+1))] \leq n + 2n^{1/2}
{( \expected[f] )}^{1/2}$.
\end{proof}

\begin{lemma}
Either there is a $t'<t$ s.t.~$\expected[\Phi(X(t'))]\leq 18n$
or
$\expected[\Phi(X(t))] \leq 9^{1-2^{-t}} n^{1-2^{-t}}
{\Phi(X(0))}^{2^{-t}}$.
\end{lemma}
\begin{proof}
The proof is by induction on $t$. The base case is $t=0$. For the inductive
step, note that $1-2^{-t}=\sum_{k=1}^t 2^{-k}$. Suppose that for all $t'< t$,
$\expected[\Phi(X(t'))]> 18n$ (otherwise we are finished). Then by
Corollary~\ref{corjensen},
\[
   \expected[\Phi(X(t))]
   =    n + 2n^{1/2} {(\expected[\Phi(X(t-1))])}^{1/2}
   \le  3n^{1/2} {(\expected[\Phi(X(t-1))])}^{1/2}.
\]
Applying the inductive hypothesis,
\[
  \expected[\Phi(X(t))] \leq
  3n^{1/2}
  {(
  3^{2(1-2^{-(t-1)})} n^{1-2^{-(t-1)}} {\Phi(X(0))}^{2^{-(t-1)}}
  )}^{1/2}.
\]
\end{proof}

\begin{corollary}\label{lowpotentialfast}
There  is a $\tau\leq \lceil \lg\lg {\Phi(X(0))} \rceil$ such that
$\expected[\Phi(X(\tau))]\leq 18n$.
\end{corollary}
\begin{proof}
Take $t=\lceil \lg\lg {\Phi(X(0))} \rceil$. Either there is a $\tau<t$ with
$\expected[\Phi(X(\tau))]\leq 18n$ or, by the lemma,
$$\expected[\Phi(X(t))] \leq 9  n  {\Phi(X(0))}^{2^{-t}}
\leq 18n.$$
\end{proof}

\begin{corollary}
\label{cortau} There is a $\tau\leq \lceil \lg\lg {\Phi(X(0))} \rceil$ such
that $\Pr(\Phi(X(\tau))>720n)\leq 1/40$.
\end{corollary}
\begin{proof}
Consider the (non-negative) random variable $Y=\Phi(X(\tau))$ where
$\tau$ is the quantity from Corollary~\ref{lowpotentialfast}.
Markov's inequality says that for any $a>0$, $\Pr(Y\geq a) \leq
\expected[Y]/a$.  Now use Corollary~\ref{lowpotentialfast} with
$a=720n$.
\end{proof}

\begin{corollary}\label{epsilonnash}
For all $\epsilon>0$, provided that $n<m^{1/3}$, the expected time
to reach $\epsilon$-Nash equilibrium is $O(\log\log m)$.
\end{corollary}

\begin{proof}
Since the  bound is  asymptotic as a function of~$m$ for
fixed~$\epsilon$, we can assume without loss of generality that
$m>(60/\epsilon)^2$
and that $\epsilon m/(2n)$ is an integer. We show that for any starting assignment
$X(0)$, there exists $\tau\leq \log\log(m^2)$ such that
$\Pr(X(\tau)~{\rm is}~\epsilon {\rm -Nash})$ $>\frac{39}{40}$. This
implies the statement of the result since the number of blocks of
$\tau$ steps needed to reach an $\epsilon$-Nash equilibrium is at
most
$$1 + \left( \frac{1}{40} \right) + \left(\frac{1}{40}\right)^2 +
\cdots = \frac{40}{39}<2.$$

Suppose assignment $x$ is not $\epsilon$-Nash. If $X(t)=x$ there
exist resources $i,j$ with $X_i(t)-X_j(t)>\epsilon m/n$.
We use the following notation. Let $\Delta = \epsilon m/(2n)$.
Let $\beta= X_i(t) - X_j(t)-2\Delta$. Note $\beta> 0$.
If $X(t+1)$
is obtained from $X(t)$ by transferring $\Delta$ tasks from
$i$ to $j$, then
\begin{align*}
\Phi(X(t))-\Phi(X(t+1)) &= X_i(t)^2 + X_j(t)^2 - X_i(t+1)^2 - X_j(t+1)^2 \\
&= (2 \Delta+\beta+X_j(t))^2 + X_j(t)^2 - (\Delta+\beta+X_j(t))^2 - (\Delta+X_j(t))^2\\
&= 2 \Delta(\Delta+\beta+X_j(t)) + \Delta^2 - \left(2 \Delta X_j(t) + \Delta^2\right)\\
&= 2 \Delta(\Delta+\beta) \geq \Delta^2 =
(\epsilon m/2n)^2.
\end{align*}
It follows that $\Phi(X(t))\geq (\epsilon m/2n)^2$.

From Corollary~\ref{cortau},
$\Pr(\Phi(X(\tau))<720n)>\frac{39}{40}$, for
$\tau=\log\log(\Phi(0))=O(\log\log m)$.

An assignment $X(\tau)$ with $\Phi(X(\tau))\leq 720n$ must be
$\epsilon$-Nash if $(\epsilon m/2n)^2 > 720n$. Note that
$m>n^3$ and $m>(60/\epsilon)^2$. Hence, from
$\epsilon^2(60/\epsilon)^2 n^3 > 4.720.n^3$, we can deduce
$\epsilon^2 m^2 > 4.720.n^3$, hence
$(\epsilon m/2n)^2 > 720n$.
\end{proof}

Corollary~\ref{cortau} tells us that $\Phi(X(\tau))$ is likely to
be~$O(n)$. We want to show that $\Phi(X(t))$ quickly gets even
smaller (all the way to a Nash equilibrium) and to this end, we show
that $\Phi(X(t))$ is a super-martingale.  By Observation~\ref{obsu},
it suffices to show $u(x)\leq \Phi(x)$, and we proceed with this. In
the following, we shall consider the cases $|x_i-\overline{x}|<2.5$
for all $i\in[n]$ (Lemma \ref{lemfive}) and $\exists i\in[n]:
|x_i-\overline{x}|\ge 2.5$ (Lemma \ref{lem:pot}) separately.

\begin{lemma}
\label{lemfive}
Suppose that assignment $x=(x_1,\ldots,x_n)$ satisfies $|x_i-\overline{x}|<
2.5$ for all $i\in[n]$. Then $u(x)\leq \Phi(x)$.
\end{lemma}

\begin{proof}
For all $i\in [n]$ and $j\in[n]$ we have
$|x_i-x_j| \leq |x_i-\overline{x}| + |x_j-\overline{x}|<5$.
Let $z=\min_i x_i$ so every $x_i\in\{z,\ldots,z+4\}$. Let $n_i = |\{j
\mid x_j = z+i \}|$. Then
\[
    n^2 \Phi(x)
  = n^2 \sum_{i=1}^n x_i^2-n {\left(\sum_{i=1}^n x_i\right)}^2
  = n^2 \left( \sum_{j=0}^4 n_j {(z+j)}^2\right) -
      {\left(\sum_{j=0}^4 n_j (z+j)\right)}^2.
\]
Also, $n^2u(x) = n u_1(x) + u_2(x)$, where
$$
u_1(x) = n_0(2n_2 + 3n_3 + 4n_4) + n_1(2n_3 + 3n_4)
       + n_2(2n_0 + 2n_4) + n_3(3n_0 + 2n_1) + n_4(4n_0 + 3n_1 + 2n_2)
$$
and
\[
u_2(x) = n_0 n_1^2 + n_1(n_0 - n_2)^2 + n_2(n_1 - n_3)^2
       + n_3(n_2 - n_4)^2 + n_4 n_3^2.
\]
Plugging in these expressions and simplifying, we get
\begin{align*}
n^2 \Phi(x) - n^2 u(x) =
& 4n_0n_1n_2 + 3n_0^2n_3 + 4n_0n_1n_3 + 4n_0n_2n_3
 + 4n_1n_2n_3 + 3n_0n_3^2 + 8n_0^2n_4 +  12n_0n_1n_4  \\
& + 3n_1^2n_4 + 8n_0n_2n_4 + 4n_1n_2n_4 + 12n_0n_3n_4
 + 4n_1n_3n_4 + 4n_2n_3n_4 + 8n_0n_4^2 + 3n_1n_4^2,
\end{align*}
\noindent which is clearly non-negative since all coefficients are positive.
\end{proof}

\begin{lemma}\label{lem:pot}
Suppose that assignment $x=(x_1,\ldots,x_n)$ satisfies
$|x_n-\overline{x}|\geq 2.5$ and, for all $i\in[n]$,
$|x_i-\overline{x}| \leq |x_n-\overline{x}|$. Let
$w=(w_1,\ldots,w_{n-1})$ be the assignment with $w_i=x_i$ for
$i\in[n-1]$. Then $\Phi(x)-u(x) \geq \Phi(w)-u(w)$, that is, the
lower bound on the  potential drop for $x$ is at least as big as
that for $w$. \label{lemreduce}
\end{lemma}

\begin{proof}
Let $k=|x_n-\overline{x}|$.
We will show
\begin{enumerate}[(1)]
\item $\Phi(x)-\Phi(w)\geq k^2$, and
\item $u(x)-u(w)\leq 2k+1$.
\end{enumerate}
Then
$$\Phi(x)-u(x)-(\Phi(w)-u(w)) \geq k^2 - (2k+1),$$
which is non-negative since $k\geq 2.5\geq 1+\sqrt{2}$.
\smallskip

\noindent
First, we prove~(1).
Let $f(z)=\sum_{i=1}^{n-1}{(x_i-z)}^2$.
Note that the derivative of $f(z)$ is
$$f'(z) = 2(n-1)z - 2 \sum_{i=1}^{n-1} x_i =
2(n-1)z - 2(n-1)\overline{w}.$$
Furthermore the second derivative is $f''(z)=2(n-1)\geq 0$.
Thus, $f(z)$ is minimized at $z=\overline{w}$.
Now note that
$$\Phi(x)-\Phi(w) = k^2 + \sum_{i=1}^{n-1}{(x_i-\overline{x})}^2 -
\sum_{i=1}^{n-1}{(x_i-\overline{w})}^2\geq k^2.$$

\noindent
Now we finish the proof by proving~(2).
Assume first that $x_n=\overline{x}+k$.
Then
$$
u_1(x)-u_1(w)
= 2\sum_{i\in[n]: |x_i-x_n|>1} |x_i-x_n| \leq
   2 \sum_{i=1}^n |x_i-x_n| \\
= 2\sum_{i=1}^n (x_n-x_i) = 2nk.
$$
Let $z_j = |\{\ell \mid x_{\ell} = j\}|$.
Clearly $z_j=0$ for $j>x_n$.
Let $\M=\lceil x_n-2k \rceil$.
For $\ell\in [n]$ we have $x_{\ell}\geq \overline{x}-k = x_n - 2k$ so
$z_j=0$ for
$j<\M$.
Now $u_2(x)=\sum_{j=\M}^{x_n} z_j{(z_{j-1}-z_{j+1})}^2$.
The representation of~$w$ in terms of $z_j$s is the same as the
representation of $x$
except that $z_{x_n}$ is reduced by one. Therefore,
\begin{align*}
u_2(x)-u_2(w)
& = z_{x_n-1}\left({(z_{x_n-2}-z_{x_n})}^2 -
    {(z_{x_n-2}-z_{x_n}+1)}^2 \right) +
  + {(z_{x_n-1}-z_{x_n+1})}^2 \\
& = z_{x_n-1}(-2 z_{x_n-2}+2 z_{x_n} + z_{x_n-1}-1)
\leq z_{x_n-1}(2 z_{x_n} + z_{x_n-1}).
\end{align*}
But since $z_{x_n} \leq n-z_{x_n-1}$,
the upper bound on the right-hand side is at most
$$z_{x_n-1}(2n-2 z_{x_n-1} + z_{x_n-1})
= 2 z_{x_n-1}(n- z_{x_n-1}/2),$$ which is at most $n^2$ since the right-hand
side is
maximized at
$z_{x_n-1}=n$.
To finish the proof of~(2), use the definition of~$u$ to deduce that
$$u(x)-u(w) \leq \frac{u_1(x)-u_1(w)}{n}+\frac{u_2(x)-u_2(w)}{n^2}.$$

\noindent
The proof of~(2) when $x_n=\overline{x}-k$ is similar.
\end{proof}

\begin{corollary}
For any assignment~$x=(x_1,\ldots,x_n)$, $\Phi(x)-u(x)\geq 0$.
\label{upperu}
\end{corollary}
\begin{proof}
The proof is by induction on~$n$.
The base case, $n=1$, follows from Lemma~\ref{lemfive}.
Suppose $n>1$.
Neither $\Phi(x)$ nor $u(x)$ depends upon the order of the components in $x$,
so assume without loss of generality that
$|x_i-\overline{x}| \leq |x_n-\overline{x}|$ for all~$i$.
If $|x_n-\overline{x}|<2.5$ then apply Lemma~\ref{lemfive}. Otherwise, use
Lemma~\ref{lemreduce} to find an assignment $w=(w_1,\ldots,w_{n-1})$
such that $\Phi(x)-u(x) \geq \Phi(w)-u(w)$.
By the inductive hypothesis, $\Phi(w)-u(w)\geq 0$.
\end{proof}

Together, Observation~\ref{obsu} and Corollary~\ref{upperu} tell us that
$\expected[\Phi(X(t+1))\mid X(t)=x] \leq \Phi(x)$.
The next lemma will be used to
give a lower bound on the variance of the process.
Let $V=0.4 n^{-2}$.

\begin{lemma}\label{lemvariance}
Suppose that $X(t)=x$ and that $x$ is not a Nash equilibrium. Then
$$\Pr(\Phi(X(t+1))\neq \Phi(x)\mid X(t)=x)\geq V.$$
\end{lemma}

\begin{proof}
Choose $s$ and $\ell$ such that for all $i\in[n]$,
$x_{s}\leq x_i \leq x_{\ell}$.
Since $x$ is not a Nash equilibrium, $x_{\ell}>x_s+1$.
Assuming $X(t)=x$, consider the following experiment for
choosing $X(t+1)$.

The intuition behind the experiment is as follows.  We wish to show
that the transition from $X(t)$ to $X(t+1)$ has some variance in the
sense that $\Phi(X(t+1))$ is sufficiently likely to differ from
$\Phi(X(t))$.  To do this, we single out a ``least loaded'' resource
$s$ and a ``most loaded'' resource $\ell$ as above. In the transition
from~$X(t)$ to~$X(t+1)$ we make transitions from resources other than
resource~$\ell$ in the usual way. We pay special attention to
transitions from resource $\ell$ (and particular attention to
transitions from resource~$\ell$ which could either go to resource~$s$
or stay at resource~$\ell$).  It helps to be very precise about how
the random decisions involving tasks that start at resource~$\ell$ are
made. In particular, for each task $b$ that starts at resource~$\ell$,
we first make a decision about whether $b$ would \emph{accept} the
transition from resource $\ell$ to resource~$s$ \emph{if $b$ happened
to choose resource~$s$}. Then we make the decision about which
resource task~$b$ should choose.  Of course, we can't cheat and we
have to sample from the original required distribution.  Here are the
details.

Independently, for every $i\neq \ell$, choose
$(Y_{i,1}(x),\ldots,Y_{i,n}(x))$ from the multinomial distribution
described in Section~\ref{sec:notation}.  (In the informal description
above, this corresponds to making transitions from resources other
than resource~$\ell$ in the usual way.)  Now, for every task $b\in
x_\ell$, let $z_b=1$ with probability $1-x_s/x_\ell$ and $z_b=0$
otherwise.  (In the informal description above, this corresponds to
deciding whether $b$ would \emph{accept} the transition to~$s$ if
resource $s$ were (later) chosen.)  Let $x_\ell^+$ be the number of
tasks~$b$ with $z_b=1$ and let $x_\ell^-$ be the number of tasks~$b$
with $z_b=0$.  Choose $(Y^+_{\ell,1}(x),\ldots,Y^+_{\ell,n}(x))$ from
a multinomial distribution with the constraint $\sum_{j=1}^n
Y^+_{\ell,j}(x) = x^+_{\ell}$ and probabilities given by
$$ p^+_{\ell,j}(x) = \left\{
\begin{array}{ll}
\frac1n & \mbox{if $j=s$,}\\
\frac1n \left(1- \frac{x_j}{x_\ell}\right) & \mbox{if  $j\neq s$ and $x_\ell>x_j+1$,}\\
0 & \mbox{if $\ell\neq j$ but $x_\ell \leq x_j+1$,}\\
1 - \sum_{j\neq \ell} p_{\ell,j}(x) & \mbox{if $\ell=j.$}
\end{array}
\right.
$$
Similarly, choose
$(Y^-_{\ell,1}(x),\ldots,Y^-_{\ell,n}(x))$
from a multinomial distribution with
the constraint $\sum_{j=1}^n Y^-_{\ell,j}(x) = x^-_{\ell}$ and
probabilities
given by
$$ p^-_{\ell,j}(x) = \left\{
\begin{array}{ll}
0 & \mbox{if $j=s$,}\\
\frac1n \left(1- \frac{x_j}{x_\ell}\right) & \mbox{if  $j\neq s$ and $x_\ell>x_j+1$,}\\
0 & \mbox{if $\ell\neq j$ but $x_\ell \leq x_j+1$,}\\
1 - \sum_{j\neq \ell} p_{\ell,j}(x) & \mbox{if $\ell=j.$}
\end{array}
\right.
$$
For all~$j$, let $Y_{\ell,j}(x) = Y^+_{\ell,j}(x)+Y^-_{\ell,j}(x)$.
Informally, the $p^+_{\ell,j}$ transition probabilities are set up so
that packets which decided that they would accept a transition to~$s$
behave appropriately and the $p^-_{\ell,j}$ transition probabilities
are set up so that packets which decided that they would \emph{not}
accept a transition to~$s$ behave appropriately.  By combining the
probabilities, we see that $X(t+1)$ is chosen from the correct
distribution in this way.

Now, consider the transition from~$x$ to~$X(t+1)$.  Condition on the
choice for $(Y_{i,1}(x),\ldots,Y_{i,n}(x))$ for all $i\neq \ell$.
Suppose $x_\ell^{+}>2$.  Condition on the choice for
$(Y^-_{\ell,1}(x),\ldots,Y^-_{\ell,n}(x))$.  Flip a coin for each of
the first $x_b^{+}-2$ tasks with $z_b=1$ to determine which of
$Y^+_{\ell,1}(x),\ldots,Y^+_{\ell,n}(x)$ the task contributes to.
Condition on these choices.  Consider the following options:
\begin{enumerate}[(1)]
\item Let $x_1$ be the resulting value of $X(t+1)$ when we
add both of the last two tasks to $Y^+_{\ell,\ell}(x)$.
\item Let $x_2$ be the resulting value of $X(t+1)$ when we add one of
the last two tasks to $Y^+_{\ell,\ell}(x)$ and the other to
$Y^+_{\ell,s}(x)$.
\item Let $x_3$ be the resulting value of $X(t+1)$ when we
add both of the last two tasks to $Y^+_{s,s}(x)$.
\end{enumerate}

Note that, given the conditioning, each of these choices occurs with
probability at least $n^{-2}$.  Also, $\Phi(x_1)$, $\Phi(x_2)$ and
$\Phi(x_3)$ are not all the same.  Thus, $\Pr(\Phi(X(t+1)\neq \Phi(x)
\mid X(t)=x, x_\ell^{+}>2)\geq n^{-2}$.  Also,
$$\Pr(x_\ell^{+}>2) =
1 -{\left(\frac{x_s}{x_\ell} \right)}^{x_\ell}-
x_\ell
\left(1-
\frac{x_s}{x_\ell}\right)
{\left(\frac{x_s}{x_\ell} \right)}^{x_\ell-1}.$$
Since the derivative with respect to~$x_s$ is negative,
this is minimized by taking $x_s$ as large as possible, namely
$x_\ell-2$, so
$\Pr(x_\ell^{+}>2)\geq 1 - 7 e^{-2} \geq 0.4$, and
the result follows.
\end{proof}

In order to finish our proof of convergence, we need
the following observation about $\Phi(x)$.

\begin{observation}
For any assignment~$x$, $\Phi(x)\leq m^2$.
Let $r = m \bmod n$.
Then $\Phi(x)\geq r(1-r/n)$, with equality if and only
if $x$ is a Nash equilibrium.
\label{obsnash}
\end{observation}

\begin{proof}
Suppose that in assignment~$x$ there are resources~$i$ and~$j$ such
that $x_i-x_j\geq 2$. Let~$x'$ be the assignment constructed
from~$x$ by transferring a task from resource~$i$ to resource~$j$.
Then
\begin{align*}
\Phi(x)-\Phi(x')
& =  x_i^2-{x'_i}^2+x_j^2-{x'_j}^2
 =  x_i^2-(x_i^2-2x_i+1)+x_j^2-(x_j^2+2x_j+1) \\
& =  2x_i-2x_j-2 = 2(x_i-x_j)-2 > 0.
\end{align*}
Now suppose that, in some assignment~$x'$, resources~$i$ and~$j$ satisfy
$x'_i\geq x'_j>0$.
Let $x$ be the assignment constructed from~$x'$ by transferring a
task from resource~$j$ to resource~$i$. Since $(x'_i+1)-(x'_j-1)\geq 2$,
the above argument gives $\Phi(x)>\Phi(x')$.
We conclude that an assignment~$x$ with maximum $\Phi(x)$
must have all of the tasks in the same resource, with $\Phi(x)=m^2$.

Furthermore, an assignment $x$ with minimum $\Phi(x)$
must have
$\vert x_i-x_j\vert \leq 1$ for all $i,j$.
In this case there must be $r$ resources with loads of
$q+1$ and $n-r$ resources with loads of $q$, where $m=qn+r$.
So
$$
\Phi(x)  =  r(q+1-\bar{x})^2 + (n-r)(q-\bar{x})^2
=  r{\left(1-\frac{r}{n}\right)}^2 + (n-r){\left(
          \frac{r}{n}\right)}^2
= r\left(1-\frac{r}{n}\right).
$$
Note that $x$ is a Nash assignment if and only if
$|x_i-x_j|\leq 1$ for all~$i$ and~$j$.
\end{proof}

\noindent
Combining Observation~\ref{obsnash} and
Corollary~\ref{cortau} we find that
there is a $\tau\leq \lceil \lg\lg {m^2} \rceil$ such
that $\Pr(\Phi(X(\tau))>720n)\leq 1/40$.
Let $B=7200n + \left\lceil \frac{m^2}{n} \right\rceil  - \frac{m^2}{n}$.
Let $t'=\tau+\lceil 10B^2/V\rceil$.

\begin{lemma}
\label{lem:3/4}
Given any starting state $X(0)=x$,
the probability that $X(t')$ is a Nash equilibrium is at least $3/4$.
\end{lemma}

\begin{proof}
The proof is based on a standard martingale argument, see~\cite{lrs}.
Suppose that $\Phi(X(\tau))\leq720n$.
Let $W_t = \Phi(X(t+\tau))-r(1-r/n)$ and let
$D_t = \min(W_t,B)$. Note that $D_0 \leq 720 n$.
Together, Observation~\ref{obsu} and Corollary~\ref{upperu} tell us that
$W_t$ is a supermartingale. This implies that $D_t$ is also a
supermartingale since
\[
       \expected[D_{t+1} \mid D_t=x<B]
  \leq \expected[W_{t+1} \mid W_t=x<B]\leq W_t=D_t,
\]
and
$$\expected[D_{t+1}\mid D_t=B] \leq B=D_t.$$
Together, Lemma~\ref{lemvariance} and Observation~\ref{obsnash}
tell us that if $x>0$,
$\Pr(W_{t+1}\neq W_t\mid W_t=x)\geq V$.
Thus, if $0<x<B$,
\begin{align*}
 \Pr(D_{t+1} \neq D_t\mid D_t=x)
&= \Pr(\min(W_{t+1},B)\neq W_t\mid W_t=x)\\
&\geq \Pr(W_{t+1}\neq W_t \wedge B\neq W_t\mid W_t=x)
= \Pr(W_{t+1}\neq W_t \mid W_t=x)\geq V.
\end{align*}
Since $D_{t+1}-D_t$ is an integer,
 $\expected[{( D_{t+1} - D_t)}^2 \mid 0<D_t<B ]
\geq V$.
Let $T$ be the first time at which either
(a) $D_t=0$ (i.e., $X(t+\tau)$ is a Nash equilibrium), or
(b) $D_t=  B$. Note that $T$ is a stopping time.
Define $Z_t=(B-D_t)^2
- V t$, and observe that $Z_{t\wedge T}$ is a sub-martingale, where $t\wedge
T$
denotes the minimum of $t$ and $T$.
Let $p$ be the probability that (a) occurs.
By the optional stopping theorem $\expected[D_T] \leq D_0$,
so $(1-p)B = \expected[D_T] \leq D_0$ and
$p \geq 1- D_0/B \geq
\tfrac{9}{10}$.
Also, by the optional stopping theorem
\[
   p{B}^2 - V \expected[T]
 = \expected[ {(B-D_T)}^2] - V \expected[T]  = \expected[Z_T] \geq Z_0 \\
 = {(B-D_0)}^2>0,
\]
so
$\expected[T] \leq pB^2/V$.
Conditioning on (a) occurring, it follows that
$\expected[T \mid D_T=0] \leq B^2/V$.
 Hence $\Pr(T
> 10B^2/V\mid D_T=0) \leq \tfrac{1}{10}$. So, if we now run
for  $10B^2/V$ steps, then the probability that we do not reach a
Nash equilibrium is at most $\tfrac{1}{40}+2 \cdot \tfrac{1}{10} <
1/4$.
\end{proof}

\noindent Now we can give the proof of Theorem~\ref{thm:main}.
\begin{proof}
Subdivide time into intervals of $t'$ steps.
The probability that the process has not reached a Nash equilibrium
before the $(j+1)$st interval is at most ${(1/4)}^{-j}$.
\end{proof}

\section{Lower Bounds}\label{sec:lower}


In this section we prove the lower-bound results stated in the
introduction.
We will use the following Chernoff bound which can be found, for example,
in~\cite{hr89}.
\smallskip
\noindent Let $N\ge 1$ and let $p_i\in[0,1]$ for $i=1,\ldots,N$. Let
$X_1, X_2,\ldots,X_N$ be independent Bernoulli random variables with
$\Pr(X_i=1)=p_i$ for $i=1,\ldots,N$ and let $X=X_1+\cdots+X_N$. Then
we have $\expected[X]=\sum_{i=1}^N p_i$ and for $0\le\epsilon\le1$,
\begin{equation}
   \Pr(X \le (1-\epsilon)\cdot \expected[X]) \le
     \exp\left(-\frac{\epsilon^2\cdot \expected[X]}{3}\right)
     \label{lebound}.
\end{equation}

The following theorem gives an exponential lower bound for the
expected convergence time of the process in
Figure~\ref{fig:protslow}.

\begin{theorem}\label{thm:neutral}
Let $X(t)$ be the process in Figure~\ref{fig:protslow} with $m=n$. Let
$X(0)$ be the assignment given by $X(0)=(n,0,\ldots,0)$. Let $T$ be
the first time at which $X(t)$ is a Nash equilibrium. Then
$\expected[T]=\exp( \Theta(\sqrt{n}))$.
\end{theorem}

\begin{proof}
For an assignment~$x$, let $n_0(x)$ denote the number of resources~$i$
with $x_i=0$. Thus, $n_0(X(0))=n-1$. The (unique) Nash equilibrium~$x$
assigns one task to each resource, so $n_0(x)=0$. Let $k=\lfloor
\sqrt{n}\rfloor$. We will show that for any assignment $x$ with
$n_0(x)\geq k$, $$\Pr(n_0(X(t))<k \mid X(t-1)=x) \leq
\exp(-\Theta(\sqrt{n})).$$ This implies the result.

Suppose $X(t-1)=x$ with $n_0(x)\geq k$. For convenience, let $n_0$
denote $n_0(x)$. Let $x'$ denote $X(t)$, and let $n_0'$ denote
$n_0(x')$.
We will show that, with
probability at least $1-\exp(-\Theta(\sqrt{n}))$,
$n_0'\geq k$.
During the course of the proof, we will assume, where necessary, that
$n$ is sufficiently large.
This is without loss of generality given the $\Theta$ notation
in the statement of the result.

\paragraph{Case 1} $\Emptynum>8k$.

Consider the protocol in Figure~\ref{fig:protslow}. Let
$U = \{b \mid x_{j_b}=0\}$. $\expected[|U|] = n_0$, so by the
Chernoff bound (Equation~(\ref{lebound})),
$
\Pr(|U| \le \lceil\frac{\Emptynum}{2}\rceil
+\lceil\frac{3\Emptynum}{8}\rceil) \le
\Pr(|U| \le \frac{8}{9}\Emptynum) =
\exp\left(-\Theta(\sqrt{n})\right).
$
Thus, $|U|\geq \lceil n_0/2 \rceil + \lceil 3 n_0 /8 \rceil$ with
probability at least $1-\exp(-\Theta(\sqrt{n}))$.  Suppose this is the
case. Partition $U$ into $U_1$ and $U_2$ with $|U_1| = \lceil
n_0/2\rceil$. Let $W = \cup_{b\in U_1} \{j_b\}$. First, suppose
$|W|\leq \frac38 n_0$. In that case
\[
    |\{j \mid x'_j>0\}| \leq n - |U_1| + \frac38 n_0
  = n- \lceil n_0/2 \rceil + \frac38 n_0 \leq n-k,
\]
so $n'_0\geq k$. Otherwise,
let $U' = \{b\in U_2 \mid j_b\in W\}$.
$$\expected[|U'|] = |U_2| \frac{|W|}{n_0} \geq \frac{9}{64}n_0> \frac{9}{8}k,
$$
so by the Chernoff bound~(\ref{lebound}),
$
\Pr(|U'| \le k) = \Pr(|U'|\le (1-\frac19)\expected[|U'|])= \exp\left(-\Theta(\sqrt{n})\right)
$, recalling that $k=\lfloor\sqrt{n}\rfloor$.
Thus $|U'|\geq k$ with
probability at least $1-\exp(-\Theta(\sqrt{n}))$, which implies $n'_0\geq k$.

\paragraph{Case 2} $k\leq \Emptynum\leq 8k$.

Consider the protocol in Figure~\ref{fig:protslow}. Let $L$ be the set
of ``loners'' defined by $L=\{i\mid x_i=1\}$ and let $\ell=|L|$. The
number of resources~$i$ with $x_i>1$ is $n-n_0-\ell$ and this is at
most half as many as the number of tasks assigned to such resources
(which is $n-\ell$), so $\ell \geq n-2 n_0$. Let $U=\{b \mid i_b\in L
\mbox{ and } x_{j_b}=0 \}$. $\expected[|U|]= \ell \frac{n_0}{n} \geq
\frac{(n-2 n_0)n_0}{n} = \Theta(\sqrt{n}) $, so by the Chernoff
bound~(\ref{lebound}),
$
\Pr(|U|\le2\lceil\frac14 \ell  \frac{n_0}{n}\rceil) \le
\Pr(|U|\le\frac23 \expected[|U|]) \le \exp\left(-\Theta(\sqrt{n})\right)
$.
Thus, $|U|\geq 2 \lceil \frac14 \ell \frac{n_0}{n}\rceil$ with
probability at least $1-\exp(-\Theta(\sqrt{n}))$.
Suppose this is the case. Let $U_1$ and $U_2$ be disjoint subsets of~$U$ of
size $\lceil \frac14 \ell \frac{n_0}{n}\rceil$. Order tasks in $U$ arbitrarily
and let $S = \{ b\in U \mid \mbox{for some $b'\in U$ with $b'<b$,
$j_{b'}=j_b$.}
 \}$. (Note that $|S|$ does not depend on the ordering.)
 Let $W = \cup_{b\in U_1} \{j_b\}$.

Note that if $|W|\leq \frac15 \ell \frac{n_0}{n}$ then $|S|\geq
\frac{1}{20} \ell \frac{n_0}{n}> \frac{n_0}{40}{\left(\frac{\ell}{n}
\right)}^2$. Otherwise, let $U'=\{b\in U_2 \mid j_b \in W\}$.
\[
\expected[|U'|] = |U_2| \frac{|W|}{n_0} \geq \frac{n_0}{20}{\left(\frac{\ell}{n}\right)}^2,
\]
so, by the Chernoff bound~(\ref{lebound}),
$
\Pr(|U'|\le \frac12 \frac{n_0}{20}\left(\frac{\ell}{n}\right)^2) \le
\exp\left(-\Theta(\sqrt{n})\right)
$
(recall that $n_0\left(\frac{\ell}{n}\right)^2 \ge n_0\left(\frac{n-2 n_0}{n}\right)^2 \ge
k\left(\frac{n-16k}{n}\right)^2 = \Theta(\sqrt{n})$), and thus
$|U'| \geq \frac{n_0}{40}{\left(\frac{\ell}{n}\right)}^2$
with probability at least $1-\exp(-\Theta(\sqrt{n}))$,
so $|S| \geq
\frac{n_0}{40}{\left(\frac{\ell}{n}\right)}^2.$

Suppose then that $|S| \geq
\frac{n_0}{40}{\left(\frac{\ell}{n}\right)}^2.$ Assuming that $n$ is
sufficiently large, $|S|\geq k/41$. Let $B_0 = \cup_{b\in U} \{j_b\}$
and $B_1 = \cup_{b\in L-U} \{i_b\}$. Note that every resource in
$B_0\cup B_1$ is used in~$x'$ for some task~$b\in L$. Thus, $|B_0 \cup
B_1| \leq \ell - |S|$. Let $R=\{i\mid x_i=0\} \cup L - B_0-B_1$. Then
$|R|\geq n_0 + \ell - (\ell-|S|) \geq n_0+|S| \geq (1+\frac{1}{41})k$.

Let $T=\{b \mid i_b \not\in L, j_b\in R\}$.
$\expected[T]=(n-\ell)\frac{|R|}{n}$ and
\[
       \Pr\left(T\geq \frac{|R|}{100}\right)
  \leq \binom{n-\ell}{\frac{|R|}{100}}
         {\left(\frac{|R|}{n}\right)}^{|R|/100}
  \leq {\left(\frac{2 n_0 e 100}{n} \right)}^{|R|/100},
\]
so with probability at least $1-\exp(-\Theta(\sqrt{n}))$, $T<|R|/100$.
In that case, $n'_0 \geq |R|(1-\frac{1}{100})\geq k$.
\end{proof}

The following theorem provides a lower bound on the expected convergence
time regardless of which of the two protocols is being used.

\begin{theorem}\label{thm:lower-loglog}
Suppose that $m$ is even.  Let $X(t)$ be the process in
Figure~\ref{fig:prot} with $n=2$. Let $X(0)$ be the assignment given
by $X(0)=(m,0)$. Let $T$ be the first time at which $X(t)$ is a Nash
equilibrium. Then $\expected[T]=\Omega(\log\log m)$.  The same result
holds for the process in Figure~\ref{fig:protslow}.
\end{theorem}

\begin{proof}
Note that both protocols have the same behaviour since $m$ is even
and, therefore, the situation $x_1=x_2+1$ cannot arise. For
concreteness, focus on the protocol in Figure~\ref{fig:prot}.

Let $y(x) = \max_i x_i - m/2$ and let $y_t=y(X(t))$ so $y_0=m/2$ and,
for a Nash equilibrium~$x$, $y(x)=0$.  We will show that for any
assignment $x$, $\Pr(y_{t+1}>y(x)^{1/10}\mid X(t)=x)\geq
1-y_t^{-1/4}$.  (There is nothing very special about the exact value
``$1/10$'' -- this value is being used as part of an explicit ``lack
of concentration'' inequality in the proof, noting that for a lower
bound we essentially want to lower-bound the variances of the load
distributions. This seems to require a somewhat ad-hoc approach, in
contrast with the usage of concentration inequalities.)

 Suppose $X(t)=x$ is an assignment with $x_1\geq x_2$.  As we have
seen in Section~\ref{sec:notation}, $Y_{1,2}(x)$ (the number of
migrations from resource $1$ to resource $2$ in the round) is a
binomial random variable
\[
  B\left(x_1,\frac12 \left(1-\frac{x_2}{x_1}\right)\right) =
  B\left(\frac{m}{2}+y_t, \frac{2y_t}{ m+2y_t } \right).
\]
In general, let $T_t$ be the number of migrations from the most-loaded
resource in $X(t)$ to the least-loaded resource and note that
the distribution of $T_t$ is $B\left(\frac{m}{2}+y_t, \frac{2y_t}{ m+2y_t }
\right)$ with mean $y_t$.
If $T_t = y_t +\ell$ or $T_t=y_t-\ell$ then $y_{t+1}=\ell$.
Thus $\Pr(y_{t+1}>y_t^{1/10}) = \Pr(|T_t-\expected[T_t]|>y_t^{1/10})$.
We continue by showing that this binomial distribution is
sufficiently ``spread out'' in the region of its mode, that we
can find an upper bound on $\Pr(y_{t+1} \leq y_t^{1/10})$.
This will lead to our lower bound on the expected time for
$(y_t)_t$ to decrease below some constant (we use the constant 16).

\begin{eqnarray*}
  \Pr(T_t=y_t)
&=& { \binom{\frac{1}{2}m+y_t}{y_t} }
                 \left( \frac{2y_t}{m+2y_t} \right)^{y_t}
                 \left( \frac{m}{m+2y_t} \right)^{\frac{1}{2}m}\\
\\
\Pr(T_t=y_t+j)
&=&{ \binom{\frac{1}{2}m+y_t}{y_t+j} }
     \Bigl( \frac{2y_t}{m+2y_t} \Bigr)^{y_t+j}
     \Bigl( \frac{m}{m+2y_t} \Bigr)^{\frac{1}{2}m-j}
\end{eqnarray*}
\noindent Suppose $j>0$.
\begin{eqnarray*}
\frac{\Pr(T_t=y_t+j)}{\Pr(T_t=y_t)}
&=& \Bigl(
     \frac{2y_t}{m+2y_t}
     \Bigr)^j
     \Bigl(
     \frac{m}{m+2y_t}
     \Bigr)^{-j}
     \Bigl(
     \frac{y_t!(\frac{1}{2}m)!}{(y_t+j)!(\frac{1}{2}m+y_t-(y_t+j))!}
     \Bigr) \\
&=&
     \Bigl(\frac{2y_t}{m}\Bigr)^j
     \Bigl(
     \prod_{\ell=1}^j\frac{\frac{1}{2}m+1-\ell}{y_t+\ell}
     \Bigr)
  =
     \Bigl(\frac{2y_t}{m}\Bigr)^j
     \Bigl(
     \prod_{\ell=1}^j\frac{m+2-2\ell}{2y_t+2\ell}
     \Bigr) \\
&>&
     \Bigl(\frac{2y_t}{m}\Bigr)^j
     \Bigl(
     \prod_{\ell=1}^j\frac{m-2j}{2y_t+2j}
     \Bigr)
  =
     \Bigl[
     \Bigl(\frac{2y_t}{m}\Bigr)
     \Bigl(\frac{m-2j}{2y_t+2j}\Bigr)
     \Bigr]^j.
\end{eqnarray*}

\noindent Similarly, for $j<0$,
\begin{eqnarray*}
\frac{\Pr(T_t=y_t+j)}{\Pr(T_t=y_t)}
&=&
     \Bigl(
     \frac{2y_t}{m}
     \Bigr)^j
     \Bigl(
     \prod_{\ell=1}^{\vert j\vert}\frac{y_t+1-\ell}{\frac{1}{2}m+\ell}
     \Bigr)
  =
     \Bigl(\frac{m}{2y_t}\Bigr)^{\vert j\vert}
     \Bigl(
     \prod_{\ell=1}^{\vert j\vert}\frac{2y_t+2-2\ell}{m+2\ell}
     \Bigr) \\
&>&
     \Bigl(\frac{m}{2y_t}\Bigr)^{\vert j\vert}
     \Bigl(
     \frac{2y_t-2\vert j\vert}{m+2\vert j\vert}
     \Bigr)^{\vert j\vert}
  =
     \Bigl[
     \Bigl(\frac{m}{2y_t}\Bigr)\Bigl(\frac{2y_t-2\vert j\vert}{m+2\vert j\vert}
     \Bigr)
     \Bigr]^{\vert j\vert} \\
&=&
     \Bigl[
     \Bigl(\frac{2y_t}{m}\Bigr)
     \Bigl(\frac{m-2j}{2y_t+2j}\Bigr)
     \Bigr]^j.
\end{eqnarray*}
\noindent So for all $j$,
\[
  \frac{\Pr(T_t=y_t+j)}{\Pr(T_t=y_t)}
  >
     \Bigl[
     \Bigl(\frac{2y_t}{m}\Bigr)
     \Bigl(\frac{m-2j}{2y_t+2j}
     \Bigr)
     \Bigr]^{j}
  =
     \Bigl[
     \Bigl(\frac{y_t}{y_t+j}\Bigr)
     \Bigl(\frac{m-2j}{m}
     \Bigr)
     \Bigr]^{j}.
\]

So, for all $j$ with $\vert j\vert\leq y_t^{1/4}$,
where $y_t^{1/4}$ is the positive fourth root of $y_t$,
this is at least
\begin{eqnarray*}
\lefteqn{\Bigl( \frac{y_t}{y_t+y_t^{1/4}} \Bigr)^{y_t^{1/4}}
         \Bigl( \frac{m-2y_t^{1/4}}{m} \Bigr)^{y_t^{1/4}}} \\
&\geq&
    \Bigl( \frac{y_t}{y_t+y_t^{1/4}} \Bigr)^{y_t^{1/4}}
    \Bigl( \frac{2y_t-2y_t^{1/4}}{2y_t} \Bigr)^{y_t^{1/4}}
=
    \Bigl( \frac{y_t-y_t^{1/4}}{y_t+y_t^{1/4}} \Bigr)^{y_t^{1/4}}
=
   \left(
    \frac{ y_t + y_t^{1/4} - 2y_t^{1/4} }{ y_t + y_t^{1/4}}
  \right)^{y_t^{1/4}} \\
&=&
  \left(
    1 - \frac{ 2y_t^{1/4} }{ y_t + y_t^{1/4} }
  \right)^{y_t^{1/4}}
\ge
  \left(
    1 - \frac{ 2y_t^{1/4} }{ y_t }
  \right)^{y_t^{1/4}}
=
  \left(
    1 - 2y_t^{-3/4}
  \right)^{y_t^{1/4}}\\
&\ge&
  1 - 2y_t^{-3/4}y_t^{1/4}
=
  1 - 2y_t^{-1/2}
  \ge
  \frac{1}{2}
\end{eqnarray*}
where the last inequality just requires $y_t\geq 16$.

Note that the mode of a binomial distribution is one or
both of the integers closest to the expectation, and the distribution
is monotonically decreasing as you move away from the mode.
But, for $\vert j\vert \leq y^{1/4}_t$,
$\Pr(T_t=y_t+j)\geq \frac{1}{2} \Pr(T_t=y_t)$,
hence $\Pr(T_t=y_t)\leq 2/(1+2 y_t^{1/4})$.
Since $\Pr(T_t=y_t+j) \leq \Pr(T_t=y_t)$, it follows that
\[
  \Pr(T_t\in [y_t-y_t^{1/10},y_t+ y_t^{1/10}])
  \leq
     (2 y_t^{1/10}+1) \Pr(T_t=y_t)
   <
      3 y_t^{-3/20}.
\]

We say that the transition from $y_t$ to $y_{t+1}$ is a
``fast round'' if
$y_{t+1} \leq y_t^{1/10}$
(equivalently, it is a fast round if $T_t \in [y_t-y_t^{1/10},y_t+
y_t^{1/10}]$).
Otherwise it is a slow round.
Recall that $y_0=m/2$.
Let
$$ r = \left\lfloor\log_{10}
\left( \frac{\log (y_0)} {\log(12^{20/3})} \right)\right\rfloor.$$ If the
first~$j$ rounds are slow then $y_j\geq y_0^{10^{-j}}$.
 If $j\leq r$ then $y_0^{10^{-j}}\geq 12^{20/3}$ so the probability
that the transition from $y_j$ to $y_{j+1}$ is the first fast round is at
most
$3{\left( y_0^{10^{-j}} \right)}^{-3/20}\leq 1/4$.

Also, if $j< r$ then these probabilities increase geometrically so
that the ratio of the probability that the transition to $y_{j+1}$ is
the first fast round and the probability that the transition to $y_j$
is the first fast round is
\[
\frac
  {3
  {
  \left(
  y_0^{
  10^{-(j+1)}
  }
  \right)
  }^{-3/20}
  }
  {3{ \left( y_0^{10^{-j}} \right)}^{-3/20}}
=
  {\left(y_0^{10^{-j}-10^{-(j+1)}}\right)}^{3/20}
\geq
  {\left(y_0^{10^{-(j+1)}}\right)}^{3/20}
  \geq 12 \geq 2,
\]
so $\sum_{j=0}^{r-1}
         \Pr(\mbox{transition~from $y_j$ to $y_{j+1}$ is the
                   first fast round)} \leq 2 \cdot 1/4 = \frac12$.
Therefore, with probability at least $1/2$, all of the first $r$
rounds are slow.  In this case, $\arg\min_t(y_t\leq
16)=\Omega(\log\log(m))$, which proves the theorem.
\end{proof}

We also have the following observation.
\begin{observation}
Let $X(t)$ be the process in Figure~\ref{fig:prot} with $m=n$. Let
$X(0)$ be the assignment given by $X(0)=(2,0,1,\ldots,1)$. Let $T$ be
the first time at which $X(t)$ is a Nash equilibrium. Then
$\expected[T]=\Omega(n)$.
\end{observation}

The observation follows from the fact that the state does not change
until one of the two tasks assigned to the first resource chooses the
second resource.

\section{Summary}

We have analyzed a very simple, strongly distributed rerouting
protocol for $m$ tasks on $n$ resources.  We have proved an upper
bound of $(\log\log m+n^4)$ on the expected convergence time
(convergence to a Nash equilibrium), and for $m>n^3$ an upper bound
of $O(\log\log m)$ on the time to reach an approximate Nash
equilibrium. Our lower bound of $\Omega(\log\log m+n)$ matches the
upper bound as function of $m$.  We have also shown an exponential
lower bound on the convergence time for a related protocol that
allows ``neutral moves''.

\end{document}